# Spatiotemporal Gaussian Optimization for 4D Cone Beam CT Reconstruction from Sparse Projections

Yabo Fu, Hao Zhang, Weixing Cai, Huiqiao Xie, Licheng Kuo, Laura Cervino, Jean Moran, Xiang Li, Tianfang Li

***Abstract*—In image-guided radiotherapy (IGRT), four-dimensional cone-beam computed tomography (4D-CBCT) is critical for assessing tumor motion during a patient's breathing cycle prior to beam delivery. However, generating 4D-CBCT images with sufficient quality requires significantly more projection images than a standard 3D-CBCT scan, leading to extended scanning times and increased imaging dose to the patient. To address these limitations, there is a strong demand for methods capable of reconstructing high-quality 4D-CBCT images from a 1-minute 3D-CBCT acquisition. The challenge lies in the sparse sampling of projections, which introduces severe streaking artifacts and compromises image quality. This paper introduces a novel framework leveraging spatiotemporal Gaussian representation for 4D-CBCT reconstruction from sparse projections, achieving a balance between streak artifact reduction, dynamic motion preservation, and fine detail restoration. Each Gaussian is characterized by its 3D position, covariance, rotation, and density. Two-dimensional X-ray projection images can be rendered from the Gaussian point cloud representation via X-ray rasterization. The properties of each Gaussian were optimized by minimizing the discrepancy between the measured projections and the rendered X-ray projections. A Gaussian deformation network is jointly optimized to deform these Gaussian properties to obtain a 4D Gaussian representation for dynamic CBCT scene modeling. The final 4D-CBCT images are reconstructed by voxelizing the 4D Gaussians, achieving a high-quality representation that preserves both motion dynamics and spatial detail. The code and reconstruction results can be found at: https://github.com/fuyabo/4DGS_for_4DCBCT/tree/main***

***Index Terms*—Cone-beam computed tomography, deep learning, reconstruction, 4D Gaussian.**

## I. INTRODUCTION

Stereotactic body radiotherapy (SBRT) requires high target localization accuracy due to its tight planning margins and high fractional dosage [1]. Four-dimensional (4D) cone-beam CT (CBCT) can be acquired to evaluate target motion before beam delivery by sorting projections into respiratory phases and reconstructing a sequence of 3D CBCT images. Despite its potential for motion assessment, clinical adoption of 4D-CBCT is hindered by long scan times, higher imaging doses and low image quality. An alternative approach is reconstructing 4D CBCT from a single 3D CBCT scan, which is routinely performed for daily patient setup. However, achieving high-quality phase-resolved images from a single conventional 1-minute scan is extremely challenging. The under-sampled phase-binned projections lead to severe streak artifacts, impairing visualization of small structures and tissue boundaries [2].

In the literature, several methods have been proposed to improve 4D-CBCT image quality, including iterative reconstruction, motion compensation, prior image deforming, and deep learning (DL)-based approaches [28]. Iterative reconstruction methods like total variation (TV) minimization focus on under-sampling problem, which suppress noise and artifacts by regularizing image gradients [3]. Under the same principle, prior image constrained compressive sensing (PICCS) further leverages the sparsity of differences between a target image and a motion-blurred prior image to reduce motion artifacts [4]. However, due to the impact of prior image, PICCS struggles to fully account for respiratory phase deformations, often leaving residual motion artifacts. Motion-compensation techniques use models built from prior image e.g. the 4DCT, before the 4D CBCT acquisition, to address under-sampling issue [5-8]. Reference [29] introduced an enhanced 4D CBCT reconstruction method utilizing three different motion modeling methods, including 4D phase-to-phase registration, CT image to 4D phases registration and CT image to CBCT projection registration for motion compensation. However, motion-compensation methods face limitations when motion patterns differ between the motion model and the 4D CBCT. Reference [8] estimates motion directly from 4D-CBCT, however, the motion model is prone to errors due to degraded intermediate CBCT images. Prior image deforming techniques [9], which rely on principal component analysis (PCA)-based motion models, also face challenges when motion patterns or anatomy differ between prior images and the images being reconstructed [9].

In recent years, deep learning has demonstrated significant

Memorial Sloan-Kettering Cancer Center has a research agreement with Varian Medical Systems. This research was partially supported by the MSK Cancer Center Support Grant/Core Grant (P30 CA008748).

The authors are with the Department of Medical Physics, Memorial Sloan Kettering Cancer Center. 1275 York Ave, New York, NY 10065. Yabo Fu is the corresponding author (e-mail: fuy3@mskcc.org).

promise in 4D-CBCT reconstruction [10-17]. CNNs such as U-Nets, residual CNNs, and DenseNets have successfully reduced artifacts and improved structural fidelity. Incorporating prior knowledge, such as average-image constraints, has further enhanced artifact removal and detail restoration, especially when combined with hybrid methods like motion correction or iterative reconstruction. However, DL methods dealing with sparse input projections are limited by poor generalization across diverse datasets, and the lack of ground-truth clinical 4D-CBCT images, as models trained on simulations often fail to generalize. Implicit neural representation (INR) learning has emerged as a novel DL technique, using multi-layer perceptrons (MLPs) to represent complex objects as continuous and differentiable functions. By querying the INR at voxel coordinates, image intensity can be reconstructed without explicitly defining constitutive functions. The Simultaneous Spatial and Temporal Implicit Neural Representation (STINR) leverages INR for dynamic CBCT reconstruction, decoupling the problem into a spatial INR for a reference CBCT image and temporal INRs for deformation vector fields (DVFs) [18-19]. However, the approach requires significant GPU memory, necessitating down-sampled matrix dimensions, which compromise image resolution. The optimization process, involving repeated voxelization of the continuous INR, also increases complexity and reconstruction time.

Recently, 3D Gaussian splatting has gained attention for efficient, high-quality novel view synthesis and surface reconstruction [20]. Initially developed for natural light imaging with spherical harmonics modeling RGB colors, its adaptation to X-ray imaging posed challenges. Cai *et al.* introduced isotropic grayscale modeling to replace spherical harmonics, enabling sparse-view CT image reconstruction through novel view synthesis, and final CT reconstruction relied on traditional methods [21]. Zha *et al.* later proposed R2-Gaussian, a radiative Gaussian splatting framework for tomographic reconstruction, replacing natural light rasterization with voxelization tailored to X-ray imaging [22]. While R2-Gaussian was designed for static 3D CT reconstruction, dynamic CBCT reconstruction demands adapting the 3D Gaussian representation to 4D to model the spatiotemporal change of the human anatomy. For natural light images, 4D Gaussian splatting has been explored for real-time dynamic scene rendering [23]. Inspired by Wu *et al.*, who integrated a deformation network into 3D Gaussian splatting for 4D dynamic view synthesis, we propose incorporating a Gaussian deformation network into the R2-Gaussian framework for dynamic CBCT reconstruction. The main contributions of this paper are as follows:

- We present a unified, end-to-end framework for dynamic CBCT reconstruction using a differentiable 4D Gaussian representation.
- The method eliminates the need for prior CT images or respiratory motion models.
- The spatiotemporal feature encoding enables shared feature learning across phase-binned projection images, effectively removing streak artifacts from under-sampled projections and accurately capturing underlying motion.

## II. PROPOSED METHOD

In this section, we first outline the overall framework of the proposed method, as illustrated in Fig. 1. The key components of the method include the 3D Gaussian splatting process and the modifications required to adapt it for X-ray imaging, as well as the Gaussian deformation field network designed for dynamic scene modeling. Finally, the proposed 4D Gaussian representation combines the 3D Gaussian splatting framework with the deformation network, which are jointly optimized to reconstruct 4D-CBCT from severely under-sampled projection images.

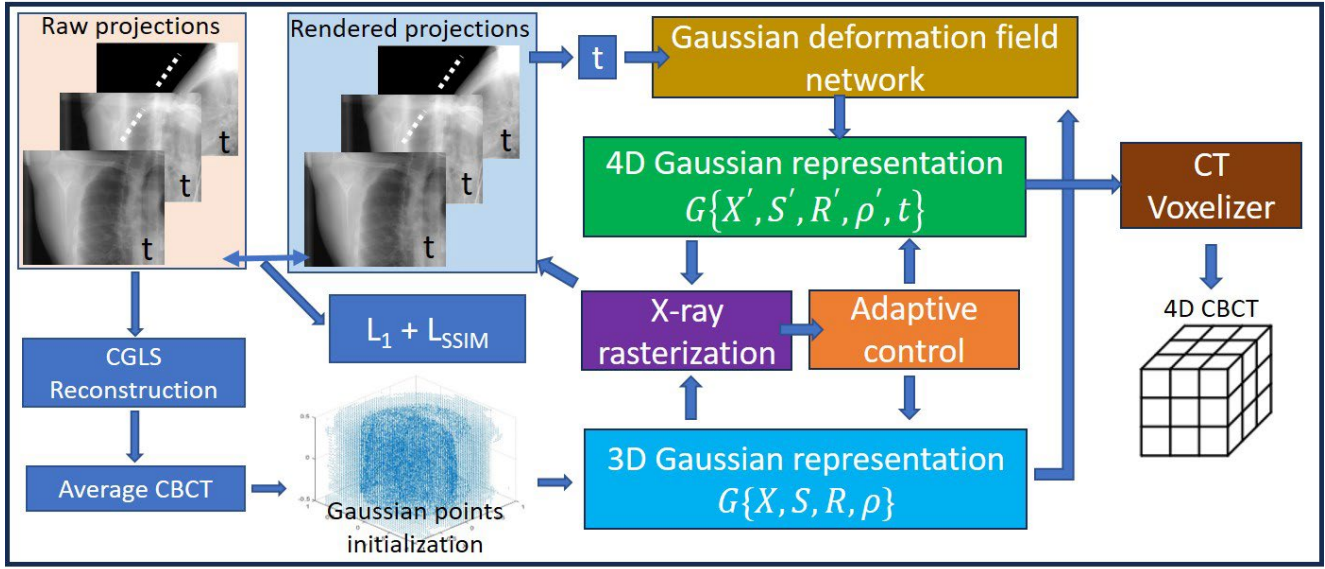

Fig. 1. The proposed 4D Gaussian representation framework for dynamic CBCT modelling.

The process begins with the reconstruction of an average CBCT using raw free-breathing CBCT projection images. This average CBCT serves as the initialization for the Gaussian points in the 3D Gaussian training. In the initial stage, the 3D Gaussians are trained to represent a static scene using all projections, under the assumption of no internal anatomical motion. This step ensures that the 3D Gaussian points can approximate the imaged object without accounting for motion.

In the subsequent stage, the trained 3D Gaussians and the timestamps of each projection image are processed through the Gaussian deformation field network. The timestamp of each projection image is the phase index which can be obtained using various phase sorting techniques [28]. This network deforms the 3D Gaussians such that the rendered projections align with the measured raw projections, considering the phase each projection belongs to. During this stage, the Gaussian deformation field network and the 3D Gaussians are jointly optimized to model a dynamic scene for the subsequent 4D-CBCT reconstruction.

In both stages, an X-ray rasterization module is employed to render the 3D Gaussian point cloud onto 2D X-ray projections, which are then compared against the raw projections to compute the loss. The total loss is a weighted combination of the L1 loss and the Structural Similarity Index Measure (SSIM) loss. An adaptive control module is integrated into both stages to densify and prune the Gaussians based on the accumulated gradients of the Gaussian properties, as well as the predefined allowable ranges for scale and density. Finally, a CT voxelizer is used to generate the reconstructed 4D-CBCT.

### A. 3D Gaussian splatting

3D Gaussian Splatting (GS) is an innovative technique used in computer graphics for rendering 3D scenes [20]. It represents scenes using explicit 3D Gaussians, structured as point clouds. Each Gaussian is characterized by a center point $X$ (mean position) and a covariance matrix $\Sigma$, which models the spatial extent and orientation of the Gaussian:

$$G(X) = e^{-\frac{1}{2}X^T \Sigma^{-1} X} \quad (1)$$

To enable differentiable optimization, $\Sigma$ is decomposed into a scaling matrix $S$ and a rotation matrix $R$:

$$\Sigma = RSS^T R^T \quad (2)$$

For rendering, the Gaussians are projected onto the camera plane using differentiable splatting. The covariance in camera coordinates, $\Sigma'$ is calculated using the view transform matrix $W$ and the Jacobian J of the projection transformation:

$$\Sigma' = JW\Sigma W^T J^T \quad (3)$$

Each Gaussian is defined by 3D position $X \in \mathbb{R}^3$, spherical harmonics (SH) coefficients $C \in \mathbb{R}^k$ representing color, opacity $\alpha$, scaling factors $s \in \mathbb{R}^3$, and rotation $r \in \mathbb{R}^4$.

The final color $C$ for a pixel is computed by blending the contributions of all overlapping Gaussians, weighted by their opacities $\alpha$:

$$C = \sum_{i \in N} c_i \alpha_i \prod_{j=1}^{i-1}(1 - \alpha_j) \quad (4)$$

Where $c_i$ and $\alpha_i$ are derived from the Gaussian properties, including its covariance matrix $\Sigma$ and SH coefficients. The 3D GS allows for efficient scene rendering while maintaining explicit control over the representation, offering flexibility for both rendering and optimization.

### B. 3D Gaussian splatting applied to X-ray

X-Ray imaging is fundamentally different from natural lighting imaging. X-ray imaging relies on density integration along rays, necessitating key modifications [22]:
- Representation: View-dependent color modeling was changed to isotropic density for accurate X-ray attenuation modeling.
- X-ray Rasterization: Density projections from 3D to 2D by integrating along ray paths. This step corrects the integration bias inherent in standard 3DGS, ensuring consistent density retrieval across views.
- Density Voxelization: A CUDA-based voxelizer efficiently retrieves volumetric density from Gaussians, enabling 3D regularization and reconstruction.

After the above modifications, the density field is represented using a set of 3D Gaussians:

$$G_i^3(x|\rho_i, p_i, \Sigma_i) = \rho_i e^{-\frac{1}{2}(x-p_i)^T \Sigma_i^{-1}(x-p_i)} \quad (5)$$

where $\rho_i$ is the central density, $p_i$ is the position, $\Sigma_i$ is the covariance of the $i^{th}$ Gaussian point. The total density at position $x$ is:

$$\sigma(x) = \sum_{i \in N} G_i^3(x|\rho_i, p_i, \Sigma_i) \quad (6)$$

For X-ray projections, the pixel value $I_r(r)$ along a ray $r(t) = o + td$ is the integral of density:

$$I_r(r) = \int \sigma(r(t)) dt \quad (7)$$

where $o$ is the x-ray source location, and $d$ is the unit vector from the source to the x-ray detector.

Substituting the radiative Gaussian definition:

$$I_r(r) = \sum_{i \in N} \int G_i^3(r(t)|\rho_i, p_i, \Sigma_i) dt \quad (8)$$

Through projection, the 3D Gaussian are transformed in to 2D Gaussians:

$$G_i^2(x|\rho_i', p_i', \Sigma_i') = \rho_i' e^{-\frac{1}{2}(x-p_i')^T \Sigma_i'^{-1}(x-p_i')} \quad (9)$$

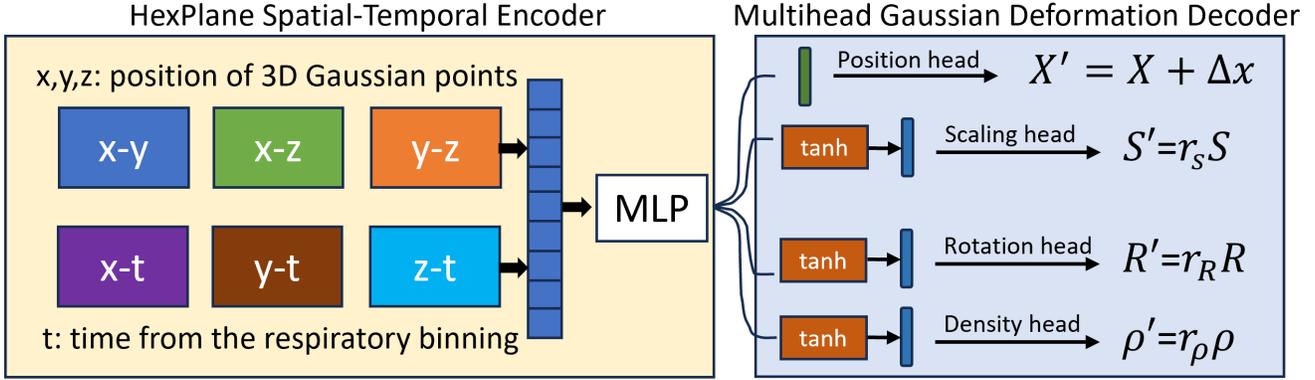

Fig. 2. The Gaussian deformation field network, consisting of the HexPlane spatial-temporal encoder and the multihead Gaussian deformation decoder.

where the density is scaled to account for the covariance factor:

$$p'_i = \rho_i \sqrt{\frac{|2\pi\Sigma_i|}{|\Sigma'_i|}} \quad (10)$$

This scaling factor corrects the integration bias present in the standard 3DGS to ensure accurate density retrieval for tomographic reconstruction [22].

### C. Gaussian deformation field network

To achieve 4D CBCT reconstruction from a sequence of dynamic projection images, the 3D Gaussians were adapted for 4D modeling. In this study, we introduce a Gaussian deformation network that leverages the powerful HexPlane spatiotemporal feature encoding [24] to capture the underlying dynamic motion from the CBCT projection image sequence. This network is designed to predict deformations for each 3D Gaussian point, encompassing changes in position, rotation, scaling, and density.

The Gaussian deformation network consists of two primary components: a spatiotemporal structure encoder and a multi-head Gaussian deformation decoder. The encoder extracts dynamic motion features from the sequence, while the decoder predicts the necessary deformations for the Gaussian points. The deformation network is jointly optimized with the 3D Gaussians using the same loss function described in Equation (14), ensuring consistency and accuracy in reconstructing the 4D CBCT.

#### 1) HexPlane spatial-temporal encoder

The HexPlane spatiotemporal encoder captures the spatial and temporal relationships among nearby Gaussians. The encoder employs multi-resolution planes, decomposing the 4D space into six 2D planes:

$$P = \{(x,y), (x,z), (y,z), (x,t), (y,t), (z,t)\} \quad (11)$$

For each plane $P_l(i,j)$, bilinear interpolation retrieves feature vectors from the voxel grid:

$$f_h = \bigcup_l \prod_{(i,j)\in P} interp(P_l(i,j)) \quad (12)$$

where the product ($\prod$) aggregates the interpolated features from all relevant planes at resolution $l$. This combines contributions from the spatial and temporal planes for a spatiotemporal representation. The union operator $\bigcup$ combines features from multiple resolutions (different levels of $l$) into a single feature vector $f_h$. Each resolution level captures details at different spatiotemporal scales, enhancing the robustness of the representation.

#### 2) Multi-head Gaussian deformation decoder

The multi-head Gaussian deformation decoder plays a crucial role in the proposed framework, enabling efficient modeling of complex dynamic scenes. This decoder operates by transforming canonical 3D Gaussians into their deformed states based on the spatial-temporal features extracted from the HexPlane encoder. The spatial-temporal features encode relevant positional and temporal information of the 4D Gaussians. The decoder $D = \{\phi_x, \phi_s, \phi_R, \phi_\rho\}$ consists of four independent multi-layer perceptron (MLPs), each dedicated to predicting specific deformation components. The positional MLP $\phi_x$ directly predict the change of each Gaussian's position. The scaling $\phi_s$, rotational $\phi_R$, and density $\phi_\rho$ MLPs were followed by a nn.tanh activation function to predict a change of ratio $r$ within range [0.9, 1.1] to avoid large changes and to stabilize the optimization. Hence the deformed Gaussians $G' = \{X', S', R', \rho'\}$ are:

$$G\{X', S', R', \rho'\} = G\{X + \Delta x, r_s S, r_R R, r_\rho \rho\} \quad (13)$$

### D. 3D Gaussian initialization

The standard 3DGS initializes Gaussian training from SfM points, which is not applicable to X-ray imaging. Instead, we initialize Gaussian training using the average CBCT images reconstructed via the traditional conjugate gradient least square (CGLS) method, as shown in Fig. 1. To avoid sampling from empty regions like air, we employ a density threshold and a CBCT field of view mask. A two-stage sampling process was proposed to ensure a proper initialization. First, a regular grid-based sampling selects one voxel out of every 8 voxels. Second, to adequately capture edges and fine structures, the gradient magnitude of the mean CBCT image is computed. From regions with gradient values exceeding a threshold, 20,000 Gaussian points are randomly sampled. Density of each Gaussian point was initialized from the CGLS reconstruction. Approximately 80,000 Gaussian points are generated per case.

This two-stage sampling balances challenges from alternative methods. Pure random sampling can lead to inaccurate CT numbers due to uneven spacing and clustered sampling, while grid-based sampling enforces uniform spacing but underrepresents fine structures. By combining these approaches, the proposed method ensures both uniform coverage and detailed structure representation, providing robust initialization for Gaussian training.

### E. 4D Gaussian optimization

To capture dynamic scenes in breathing anatomy, the training is divided into "coarse" and "fine" stages. In the coarse stage, only 3D Gaussian points are optimized to represent the average CBCT under the assumption of a static scene. After 8,000 iterations, the Gaussian deformation field network is introduced. This network deforms Gaussian points in terms of position, scaling, rotation, and density. Projection image timestamps can be determined using phase binning algorithms, such as external surrogate traces (e.g., RPM box) or projection-based methods like the Amsterdam Shroud (AS) approach. [25]. The projection images were binned into 10 phases; thus, the timestamps were set to have 10 values, ranging from 0 to 0.9 at 0.1 interval.

The 4D Gaussians were optimized using stochastic gradient descent. The Loss function is as follows:

$$L_{total} = L_1(I_m^t, I_s^t) + \lambda L_{ssim}(I_m^t, I_s^t) \qquad (14)$$

where $I_m^t$ is the measured CBCT projection images for a phase t, $I_s^t$ is the rendered projection images via X-ray rasterization from the deformed 3D Gaussian representation at timestamp of t.

Adaptive control is applied during training to improve object representation. Empty Gaussians are removed, while those with large loss gradients are densified through cloning or splitting [20]. During densification, the densities of both the original and replicated Gaussians are halved. This approach minimizes sudden performance drops caused by the introduction of new Gaussians, ensuring stable and consistent training. The densification and pruning were enabled from iteration 500 to 15000 at 100 iteration intervals. The optimization is set to terminate at iteration 30000. Starting at around 80k Gaussian points, the final number of Gaussian points are approximately 350k after adaptive control during the optimization.

In the clinic, half-fan full trajectory CBCT projection geometry is usually utilized to cover the whole patient body. The detector is shifted off-axis to achieve a larger coverage area. To accommodate the shifted detector, the FOV as defined in the pinhole cameral model is first expanded to cover the shifted detector and then cropped to match the detector's location.

### III. EXPERIMENTS AND ANALYSIS

#### A. Datasets

Public datasets from the American Association of Physicists in Medicine (AAPM) Sparse-View Reconstruction Challenge (SPARE) for 4D Cone-Beam CT (4D-CBCT) were utilized in this study. The organizers employed Monte Carlo simulations to generate clinically realistic CBCT projection datasets from lung 4DCT scans. The phase for each projection was provided. As shown in Table I, three types of simulated projection datasets were provided: low-dose (LD), scatter-free (NS), and normal scatter (SC), allowing a comprehensive benchmarking of reconstruction methods under varying scatter and imaging conditions.

TABLE I
THE THREE TYPES OF THE MONTE-CARLO SIMULATED DATASET.

|    | TUBE CURRENT | PULSE LENGTH | PRIMARY SIGNAL | QUANTUM NOISE | SCATTER NOISE |
|----|----|----|----|----|----|
| LD | 20mA | 20ms | Yes | Yes | Yes |
| NS | 40mA | 20ms | Yes | Yes | No |
| SC | 40mA | 20ms | Yes | Yes | Yes |

The public datasets provided not only the projection data with phase-binning information for reconstruction but also the ground truth data, reconstruction results for the mentioned comparison methods, and evaluation scripts. This comprehensive package is essential for ensuring a fair comparison.

As acknowledged by the AAPM SPARE challenge organizers in Reference [2], the provided clinical datasets, generated using an equi-spaced down-sampling strategy, do not accurately replicate the challenges of a real 1-minute CBCT scan. In actual clinical scenarios, projections within each respiratory bin tend to cluster around specific angles. Therefore, instead of using the provided clinical datasets, we evaluated our method using a real 1-minute scan collected in our clinic with the Varian TrueBeam onboard imager.

#### B. Comparison methods

Six methods, including the baseline Feldkamp-Davis-Kress (FDK) and five competing approaches, were used to evaluate the performance of the proposed method. The competing methods are as follows:

- **MC-FDK** (Motion-Compensated FDK): A motion-compensated reconstruction method where a prior deformation vector field (DVF) is computed from pretreatment 4DCT. This DVF is used to deform the back-projected traces, correcting for respiratory motion during reconstruction [5].

- **MA-ROOSTER** (Motion-Aware Reconstruction with Spatial and Temporal Regularization): This method utilizes a prior DVF derived from pretreatment 4DCT. Iterative reconstruction is performed to enforce spatial and temporal smoothness along a warped trajectory based on the prior DVF [26].
- **MoCo** (Motion-Compensated Data-Driven Method): A data-driven approach where the motion-compensation DVF is computed using groupwise deformable image registration of an initial 4D-CBCT reconstruction obtained through the PICCS (Prior Image Constrained Compressed Sensing) method [8].
- **MC-PICCS** (Motion-Compensated Prior Image Constrained Compressed Sensing): A variation of the PICCS method that incorporates the MC-FDK reconstruction as the prior image [27].
- **Prior Deforming**: In this method, the 4D-CBCT is generated by deforming the pretreatment 4DCT to match the CBCT projection images [9].

### C. Quantitative evaluation

The lung 4DCT scans served as the ground truth for quantitative evaluation of reconstruction performance. Image similarity was evaluated using two metrics: the root-mean-square error (RMSE) and the structural similarity index (SSIM). RMSE quantifies the pixel-by-pixel intensity differences between the reconstruction and the ground truth, with lower values indicating greater similarity. SSIM, ranging from 0 to 1, mimics human perception of image similarity by leveraging the covariance structure of pixel neighborhoods, where higher values indicate higher similarity.

For objective evaluation and comparison, all analyses, including RMSE, SSIM, and geometric assessments, were performed using the MATLAB code provided by the organizers of the AAPM SPARE challenge. The reconstruction results of these comparison methods were included in the public datasets. Four distinct regions of interest (ROIs) were considered: the patient body, lungs, planning target volume (PTV), and bony anatomy. Evaluating different ROIs provides a comprehensive assessment of an algorithm's performance, accounting for scenarios where the algorithm may produce high-quality images but fail to accurately resolve motion. In such cases, similarity metrics may yield high values in static regions (body and bony anatomy) but lower values in dynamic structures (lungs and PTV). Pixels outside the reconstruction field of view (FOV) were excluded from the ROI analysis. To comprehensively evaluate the proposed method's performance, a total of 29 simulated scans from the validation cohort were analyzed.

#### 1) RMSE and SSIM

Figure 3 summarizes the RMSE and SSIM values of the six methods for all cases in the simulated dataset. Without using any prior image, our method demonstrated comparable performance across all four examined ROIs.

Figure 4 provides a detailed comparison of RMSE_Body and SSIM_Body for all six methods, grouped by different simulation types to evaluate performance under varying scatter and noise conditions. As expected, the scatter-free dataset exhibited better overall performance compared to the other two simulation types. Within each simulation type, our method consistently achieved lower RMSE values than the competing methods, except for the prior-deforming method. However, the prior-deforming method displayed the largest standard deviations, indicating that its performance is highly sensitive to the degree of anatomical consistency between the CT and CBCT for each patient.

In terms of SSIM_Body, our method achieved intermediate performance comparable to the other five methods across all simulation types, highlighting a balanced approach to image

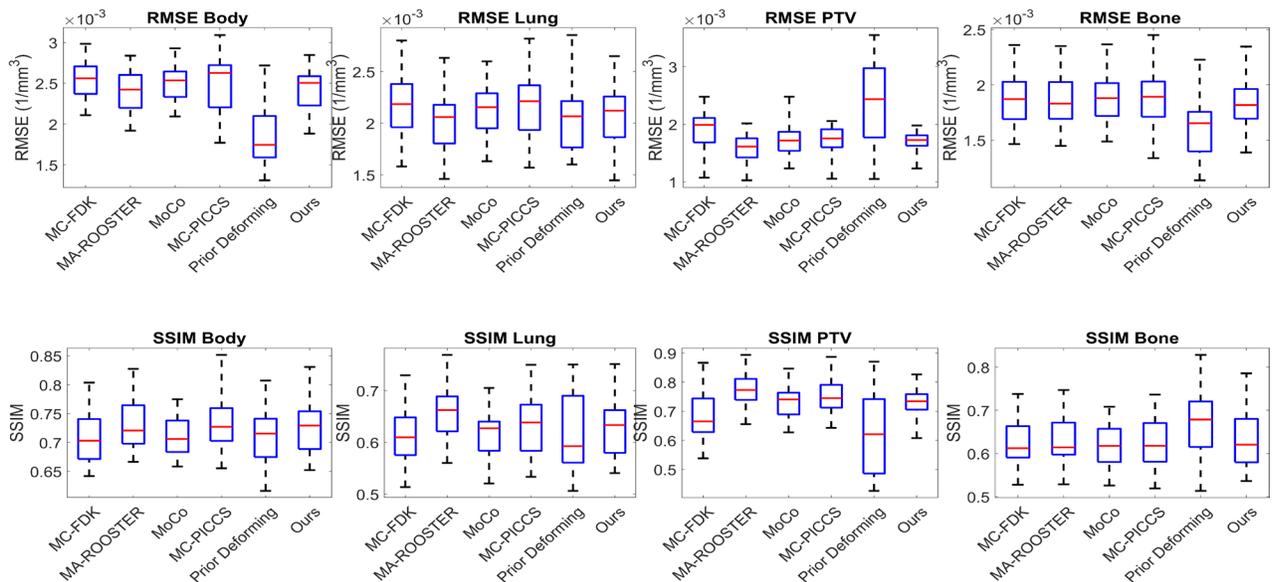

Fig. 3. Boxplots of the root-mean-squared-error (RMSE) and the structural similarity index (SSIM) in the four region-of-interest: Body, Lung, PTV and Bone for the six methods.

quality and robustness under different imaging conditions

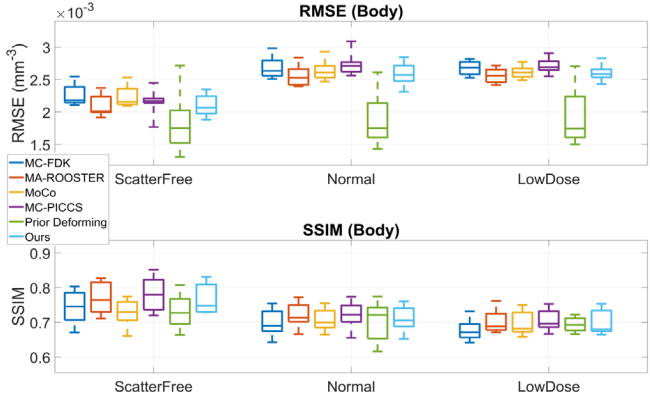

Fig. 4. The RMSE and SSIM for the body region when applied to different simulation types: Scatter-Free, Normal and Low Dose.

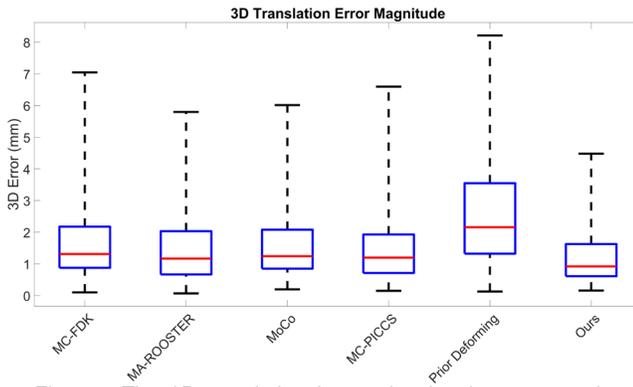

Fig. 5. The 3D translational error in planning target volume position for the six methods when applied to the simulated datasets.

### 2) Geometric accuracy

The primary clinical applications of 4D-CBCT are target alignment and verification of target motion, emphasizing the importance of geometric accuracy in the reconstructed region near the target. For geometric analysis, each phase of the reconstructed CBCT was registered to the ground truth using the Elastix package, focusing exclusively on pixels within the planning target volume (PTV). The resulting translation and rotation components were interpreted as translation and rotation errors.

Table II summarizes the translation and rotation errors for the PTV across all cases in the simulated dataset, with the best values highlighted in bold. Our method achieved the lowest errors in the LR (left-right), SI (superior-inferior), rAP (rotational anterior-posterior), and 3D magnitude of translational error. Fig. 6 further decomposes the translational error into its systemic and random components, represented by the mean and standard deviation of the error across the ten-phase reconstruction.

Except for a slightly larger systemic error in the AP (anterior posterior) direction, our method demonstrated the lowest systemic and random errors among all methods. This indicates superior accuracy and robustness in motion modeling for both

TABLE II
THE ROOT-MEAN-SQUARE(RMS) VALUES OF THE TRANSLATION AND ROTATION ERROR IN PLANNING TARGET VOLUME FOR THE SIX METHODS WHEN APPLIED TO THE SIMULATED DATASET.

|     | MC-FDK | MA-ROOSTER | MoCo  | MC-PICCS | Prior deform | Ours  |
|-----|--------|------------|-------|----------|--------------|-------|
| LR  | 0.68   | 0.71       | 0.77  | 0.95     | 1.33         | **0.54** |
| SI  | 1.47   | 1.17       | 1.22  | 0.93     | 2.50         | **0.76** |
| AP  | 1.17   | 1.15       | 1.24  | 1.25     | **0.99**     | 1.36  |
| 3D  | 2.00   | 1.79       | 1.90  | 1.82     | 3.00         | **1.65** |
| rLR | 1.35°  | 1.31°      | 1.29° | **1.28°**| 1.62°        | 1.31° |
| rSI | 0.82°  | **0.72°**  | 0.89° | 0.87°    | 1.30°        | 0.93° |
| rAP | 0.92°  | 0.72°      | 0.81° | 0.93°    | 1.19°        | **0.55°** |

[rLR, rSI, rAP] represent rotation around the LR, SI, and AP axis, respectively. Best values are shown in bold.

intra-patient and inter-patient analysis. The slightly larger systemic error in the AP direction can be attributed to Patient 6, whose alignment errors calculated by Elastix were notably higher than those of other patients across all methods, likely due to uncertainty in the Elastix registration process.

### D. Qualitative evaluation

Figures 7 and 8 illustrate the best and worst-performing cases of our method based on SSIM_Body. Like the other five competing methods, our approach significantly reduced noise and streaking artifacts compared to the FDK reconstruction. Despite the absence of explicit motion compensation, our method achieved an effective balance between image quality and geometric accuracy. As shown in Figure 8, even in the worst-performing case, our method exhibited less noticeable blurring around the diaphragm compared to the MC-FDK method.

The MC-FDK, MC-PICCS, MA-ROOSTER and Prior Deforming methods all required prior 4DCT for respiratory motion modeling and compensation. As a result, these methods demonstrated varying degrees of hallucination, especially in cases involving significant anatomical changes between CT and CBCT, as illustrated in Figure 9. For MC-FDK, MA-ROOSTER, and MC-PICCS, the tumor appeared more inferior compared to the ground truth, while for the MoCo method, it appeared more superior. Notably, the prior-deforming method, while producing the most "CT-like" image quality, failed to reconstruct the tumor region accurately due to these large anatomical changes. Although the MoCo method does not require prior images, it showed susceptibility to positional inaccuracies because its deformation vector fields (DVF) were calculated from low-quality intermediate CBCT images. In contrast, our method demonstrated more consistent performance in maintaining tumor position accuracy, evidenced by the lower 3D translational error in the planning target volume position.

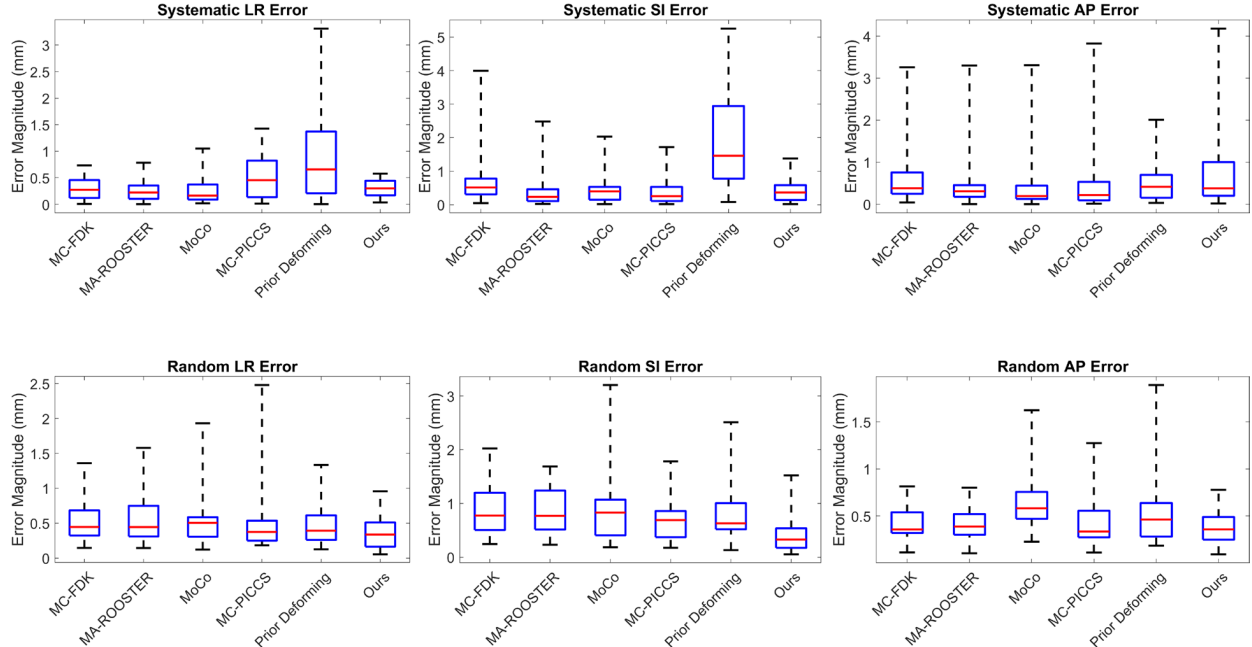

Fig. 6. The systematic and random error was calculated as the mean and standard deviation of the translation error over the ten phases of a four-dimensional cone-beam computed tomography reconstruction.

### E. Reconstruction on clinical datasets

To demonstrate the applicability of our method in a real clinical scenario, we performed 4D CBCT image reconstruction using a 1-minute CBCT scan from a Varian TrueBeam system for a lung SBRT patient. The imaging mode 'Bin2x8LowGainContinuous' was employed to acquire 893 projections in a half-fan configuration after a full rotation around the patient. The kV source voltage was set to 125 kV, with a source current of 15 mA and a pulse length of 10 ms. Projection images were preprocessed using Varian's iTools software, and phase sorted into 10 bins using the Amsterdam shroud method.

The 4D Gaussian optimization began with 50,000 Gaussian points and adaptively increased to 130,000 points through the control module by the final iteration. Despite the limited 1-minute scan duration, our reconstruction successfully suppressed streaking artifacts caused by undersampling, preserved respiratory motion, and restored fine structural details. Figure 10 illustrates the reconstructed 4D CBCT's end-inhalation and end-exhalation phases alongside the clinically used 3D FDK reconstructions. The diaphragm locations at the end-inhalation and end-exhalation phases correspond well with the motion-blurred 3D FDK reconstructions.

To evaluate the performance of the proposed deformation network in modeling high temporal motion, we sorted the projection images into 50 distinct phases instead of 10 phases. Each phase consisted of approximately 18 projections derived from 14 respiratory cycles. Using this approach, we successfully reconstructed a 4D-CBCT dataset with 50 phases, showcasing the deformation network's remarkable ability to model high-temporal dynamic motion accurately. Detailed results and supporting data are available in the accompanying GitHub repository.

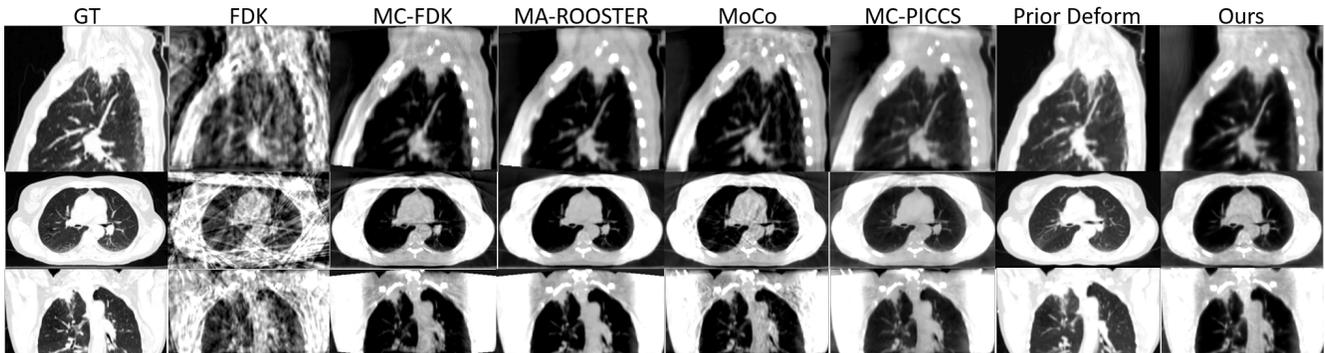

Fig. 7. The end-exhalation phase of the ground truth (GT), conventional FDK, and reconstruction from the six methods for a case on which our method has the highest SSIM Body value averaged over the 10 phases. The window level was adjusted for each method to encompass the 5[th] and 95[th] percentile pixel intensities.

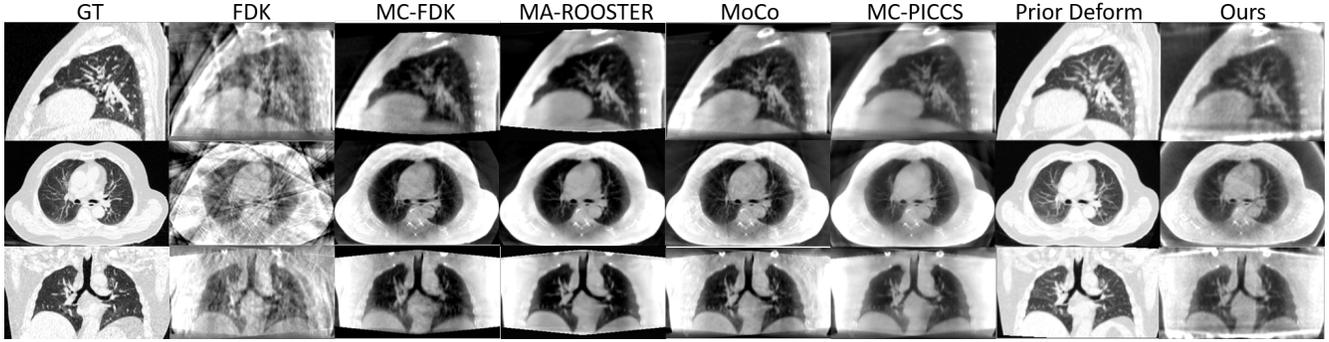

Fig. 8. The end-exhalation phase of the ground truth (GT), conventional FDK, and reconstruction from the six methods for a case on which our method has the lowest SSIM Body value averaged over the 10 phases. The window level was adjusted for each method to encompass the 5$^{th}$ and 95$^{th}$ percentile pixel intensities.

## IV. DISCUSSION

Pretreatment 4D-CBCT reconstruction from a 1-minute scan is a valuable but highly challenging task due to the severely under-sampled projections in each phase bin. Previous methods rely on building a patient-specific motion model for motion compensation. These motion models are typically represented explicitly by deformation vector fields (DVFs) to compensate for anatomic motion during CBCT acquisition. However, the registration process often introduces uncertainties to the motion modeling and limits the accuracy of the subsequent 4D CBCT reconstruction [28]. Hence, motion compensation via prior image-based motion modeling often introduces biases and fails to fully exploit the spatiotemporal signals embedded in the projection images.

This paper introduces a novel 4D-CBCT reconstruction framework leveraging the power of differentiable 4D Gaussian representations. Unlike previous approaches, the proposed method enables end-to-end modeling of the dynamic scene without requiring prior 4DCT or explicit motion compensation. The Gaussian representation of volumetric data is a relatively new concept that offers significant advantages. Given the ill-posed nature of the problem, traditional methods either reduce dimensionality e.g., by solving for coarse grid voxels, or introduce prior knowledge to compensate for missing information, or rely on regularizations to impose sparsity. However, coarse grids lose fine details, while prior knowledge can introduce artifacts and bias the reconstruction. Conversely, 3D Gaussian point clouds representation effectively reduces dimensionality by naturally adapting to the underlying structure, clustering unevenly in 3D space. For example, Gaussian points can be densely packed and tiny in one location while sparse and large in another, depending on the structural richness.

The AAPM SPARE challenge represents 4D-CBCT using a grid size of 450×450×220×10 with a 1 mm³ voxel size, equivalent to 450 million voxels/unknowns to be solved. In comparison, our method represents the 4D-CBCT using approximately 350,000 points and a Gaussian deformation network with fewer than 100,000 trainable parameters. As derived in Equation 5, each Gaussian point has 11 unknowns: 3D location, three scaling parameters, four rotation parameters, and one density parameter. This approach represents 4D-CBCT with fewer than 4 million parameters, a 100+ times reduction in the unknowns, compared to the traditional grid-based representation. The Gaussian deformation network's capability to capture underlying motion is evidenced by the lowest PTV

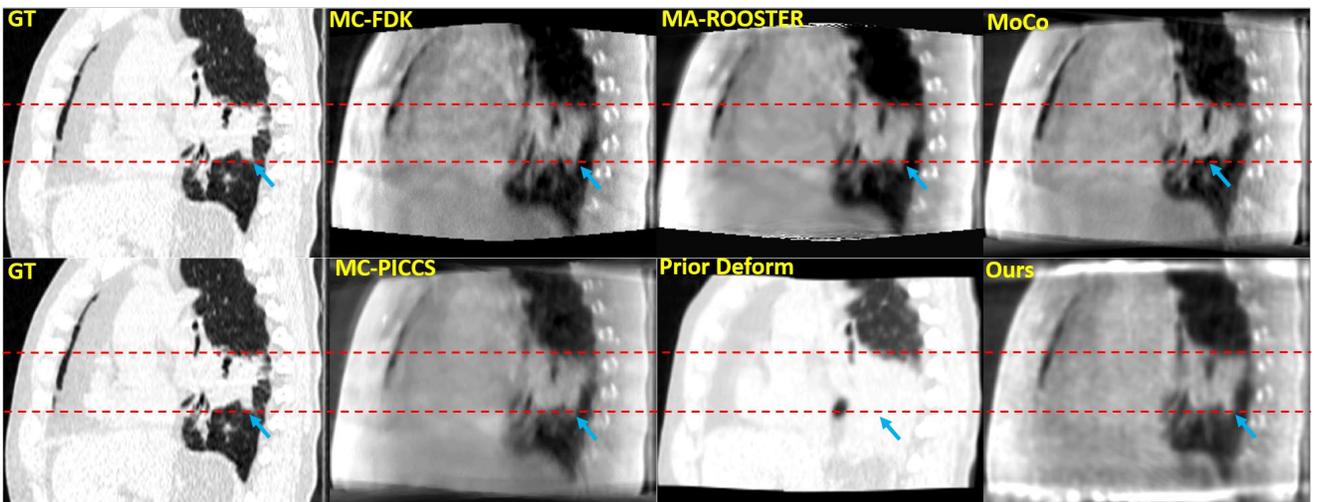

Fig. 9. The end-exhalation phase of the ground truth (GT), and reconstruction from the six methods for a case with large anatomic change between CT and CBCT. The window level was adjusted for each method to encompass the 5$^{th}$ and 95$^{th}$ percentile pixel intensities.

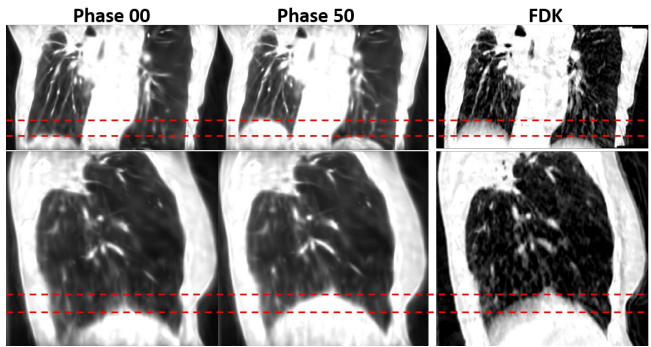

Fig. 10. Reconstruction using 1-min scan for a clinical case. Phase 00: The end-inhalation. Phase 50: the end-exhalation of the reconstructed 4D-CBCT. The last column shows the 3D CBCT reconstructed by FDK.

translation error, as shown in Figures 5 and 6.

The reconstruction results were obtained with fixed hyperparameters, without any patient- or image-specific fine-tuning, demonstrating generalizability and robustness. Consistent with the challenge guidelines, the projection images were used without scatter correction, and reconstructed images were analyzed without post-processing for objective comparison.

A limitation of the current study is the reconstruction time of approximately 3 hours for 30,000 iterations on a single NVIDIA RTX A6000 GPU, making it unsuitable for routine clinical workflows. The optimization process could be accelerated through further code improvements, such as rendering an asymmetric field of view (FOV) directly for detector shift rather than rendering a larger symmetric FOV and cropping it to the detector region or reconstructing a smaller region of interest near the tumor. Comprehensive ablation studies were not conducted to determine the minimal acceptable Gaussian points and deformation parameters (e.g., deformable vs. fixed density, scaling, and rotation). Such studies could further reduce the number of unknowns to be optimized and expedite the process.

For future work, we plan to explore regularization techniques for the Gaussian deformation network. Although our 4D Gaussian representation accurately captures dynamic CBCT scene, it is not regularized to support physics-based deformation. As a result, each deformed Gaussian does not necessarily represent realistic soft tissue deformation, limiting the ability to track Gaussian points or extrapolate/interpolate motion to time steps beyond the sorted phases. With proper physics-based regularization, we believe that 4D Gaussian representations could enable true 4D-CBCT reconstruction without the need of phase sorting, real-time motion tracking, and realistic soft tissue simulation.

ACKNOWLEDGMENT

The authors would like to thank the organizer of AAPM SPARE challenge for providing the public datasets, the participant's reconstruction results and the evaluation code, which greatly facilitated the evaluation of our method.